\documentclass[pre,twocolumn,10pt,aps]{revtex4-1}
 \usepackage{verbatim}
 \usepackage{amsmath}
 \usepackage{amssymb}
 \usepackage{amsthm}
 \usepackage{latexsym}
 \usepackage{amsfonts}
 \usepackage{epsfig}
 \usepackage{epstopdf}
 \usepackage{wasysym} 
 \usepackage{color}
 \definecolor{darkblue}{rgb}{0,0,.5}
 \usepackage[linktocpage, colorlinks=true, linkcolor=darkblue, citecolor=darkblue]{hyperref}
 \usepackage[all]{hypcap}
 \usepackage[makeroom]{cancel}

\newcommand{\C}[1]{{\cal{#1}}}
\newcommand{\bb}[1]{\textbf{#1}}

\newcommand{\lr}[1]{{\left\langle {#1}\right\rangle}}

\begin{document}

\title{Repeated interactions and quantum stochastic thermodynamics at strong coupling}

\author{Philipp Strasberg}
\affiliation{F\'isica Te\`orica: Informaci\'o i Fen\`omens Qu\`antics, Departament de F\'isica, Universitat Aut\`onoma de Barcelona, 08193 Bellaterra (Barcelona), Spain}

\date{\today}

\begin{abstract}
 The thermodynamic framework of repeated interactions is generalized to an arbitrary open quantum system in contact with 
 a heat bath. Based on these findings the theory is then extended to arbitrary measurements performed on the system. This 
 constitutes a direct experimentally testable framework in strong coupling quantum thermodynamics. By construction, it 
 provides many quantum stochastic processes and quantum causal models with a consistent thermodynamic interpretation. 
 The setting can be further used, for instance, to rigorously investigate the interplay between non-Markovianity and 
 nonequilibrium thermodynamics. 
\end{abstract}

\maketitle

\newtheorem{mydef}{Definition}[section]
\newtheorem{lemma}{Lemma}[section]
\newtheorem{thm}{Theorem}[section]
\newtheorem{crllr}{Corollary}[section]
\newtheorem*{thm*}{Theorem}
\theoremstyle{remark}
\newtheorem{rmrk}{Remark}[section]

\emph{Introduction.---} Formulating the laws of quantum thermodynamics forces us to rethink many assumptions, which are 
traditionally taken for granted. In particular, small systems are dominated by fluctuations and in general they do not 
interact weakly with a Markovian heat bath. Also the desire to monitor and manipulate quantum systems adds 
another layer of complexity due to the non-trivial effect of quantum measurements. 

In this Letter we present a unified thermodynamic framework, which overcomes the assumption of a weakly coupled, 
Markovian heat bath and which allows to include nonequilibrium resources and quantum measurements. These nonequilibrium 
resources are a set of small, externally prepared systems -- called `units' in the following -- which are sequentially 
put into contact with the system under study. This setup is known as the `repeated interaction framework' or 
`collisional model' and it has recently attracted much attention in quantum 
thermodynamics~\cite{BruneauJoyeMerkliAHP2010, HorowitzPRE2012, HorowitzParrondoNJP2013, BarraSciRep2015, 
UzdinLevyKosloffEntropy2016, PezzuttoPaternostroOmarNJP2016, StrasbergEtAlPRX2017, BenoistEtAlCM2018, 
ManzanoHorowitzParrondoPRX2018, DeChiaraEtAlNJP2018, CresserPS2019, SeahNimmrichterScaraniPRE2019, 
BaeumerEtAlQuantum2019}. However, the coupling to an additional external heat bath (typically present in an experiment) 
was mostly ignored, a weakly coupled Markovian one was only treated in Refs.~\cite{BruneauJoyeMerkliAHP2010, 
StrasbergEtAlPRX2017, CresserPS2019}. Based on recent progress in strong coupling thermodynamics~\cite{SeifertPRL2016, 
StrasbergEspositoPRE2019}, we will show that even the assumption of a weakly coupled macroscopic heat bath can be 
completely overcome. 

Afterwards, following the operational approach to quantum stochastic thermodynamics~\cite{StrasbergPRE2019, 
StrasbergWinterPRE2019}, we will show how to explicitly take into account measurements into the thermodynamic 
description. This constitutes a crucial step in strong coupling quantum thermodynamics where different strategies were 
used to arrive at many interesting conclusions~\cite{CampisiTalknerHaenggiPRL2009, 
EspositoLindenbergVandenBroeckNJP2010, TakaraHasegawaDriebePLA2010, SchallerEtAlNJP2013, GallegoRieraEisertNJP2014, 
EspositoOchoaGalperinPRL2015, EspositoOchoaGalperinPRB2015, GelbwaserKlimovskyAspuruGuzikJPCL2015, StrasbergEtAlNJP2016, 
BruchEtAlPRB2016, KatzKosloffEnt2016, LudovicoEtAlEntropy2016, NewmanMintertNazirPRE2017, MuEtAlNJP2017, 
LudovicoEtAlPRB2018, SchallerEtAlPRB2018, StrasbergEtAlPRB2018, PerarnauLlobetEtAlPRL2018, 
BruchLewenkopfVonOppenPRL2018, WhitneyPRB2018, RestrepoEtAlNJP2018, DouEtAlPRB2018, SchallerNazirBook2018, 
GuarnieriEtAlPRR2019}. However, all strategies rely on a formalism without any explicit measurements, thus making 
them hard to test and compare~\footnote{Some exceptions rely on projective measurements of the bath degrees of 
freedom~\cite{CampisiTalknerHaenggiPRL2009, SchallerEtAlNJP2013}. There is currently no technology which allows to 
carry out such measurements. Another notable exception is Ref.~\cite{LudovicoEtAlPRB2018}. }. In contrast, our theory 
is in principle immediately testable in a lab as it only requires to measure the system. Finally, we rigorously connect 
our thermodynamic framework to the field of quantum non-Markovianity. 

\begin{figure}[t]
 \centering\includegraphics[width=0.42\textwidth,clip=true]{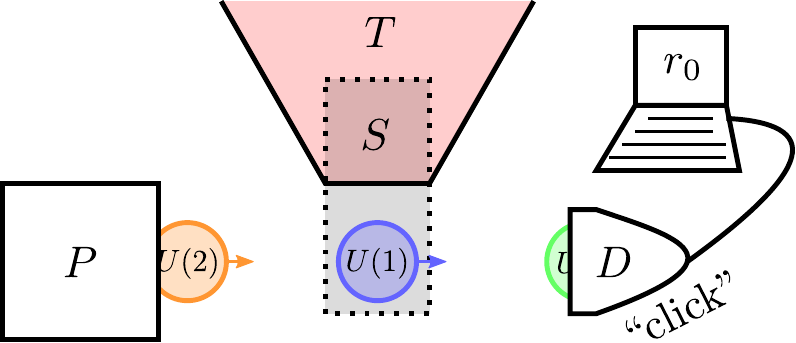}
 \label{fig setup} 
 \caption{A system $S$ is in contact with a bath initialized at temperature $T$ (we do not imply that the bath must be 
 kept at a well-defined temperature at later times). In a preparation apparatus $P$ units $U(k)$ are sequentially 
 produced, which interact with the system when they enter the shaded grey area, and which are afterwards detected in 
 $D$ giving rise to the measurement outcome $r_k$ (properly introduced later on in the text). Note that each unit can 
 be different in principle. We also remark that the description of the units does not need to be taken literally in the 
 sense that they are merely `ancillas' required for a consistent thermodynamic description of a non-Markovian quantum 
 stochastic process as introduced later on. }
\end{figure}

\emph{Setting.---} We start by considering a system $S$ coupled to a bath $B$ described by the Hamiltonian 
$H_{SB}(\lambda_t) = H_S(\lambda_t) + V_{SB} + H_B$, where $\lambda_t$ denotes an externally specified driving protocol 
(e.g., a laser field) and $V_{SB}$ denotes the system-bath interaction Hamiltonian. To this setup we add the framework 
of repeated interactions specified by the following global Hamiltonian: 
\begin{equation}\label{eq Hamiltonian}
 H_\text{tot}(\lambda_t) = H_{SB}(\lambda_t) + \sum_{k=0}^n V_{SU(k)}(\lambda_t).
\end{equation}
Here, $V_{SU(k)}(\lambda_t)$ describes the time-dependent coupling between the system and unit $U(k)$, 
$k\in\{0,\dots,n\}$, which is designed in such a way that at most one unit interacts with the system at a given time. 
Specifically, if we denote the interaction interval between the system and the $k$'th unit by $I_k \equiv [t_k,t_{k+1})$, 
then $V_{SU(k)}(\lambda_t) = 0$ for all $t\notin I_k$. Within $I_k$ the time-dependence as specified by $\lambda_t$ is 
arbitrary. Furthermore, we temporarily assume the bare unit Hamiltonian to be degenerate, i.e., $H_{U(k)} \sim 1_{U(k)}$. 
The problem is completely specified by fixing the global initial state, which is assumed to be of the form 
\begin{equation}\label{eq initial state}
 \rho_\text{tot}(t_0^-) = \pi_{SB}(\lambda_0) \otimes \rho_{U(0)}\otimes\dots\otimes\rho_{U(n)}.
\end{equation}
Here and in general we use the notation $t^\pm$ to denote the time $t\pm\epsilon$ in the limit where $\epsilon>0$ 
becomes immeasurably small. Furthermore, $\pi_X = e^{-\beta H_X}/\C Z_X$ denotes the equilibrium Gibbs state of some 
system $X$ at inverse temperature $\beta$ (perhaps depending on the value of some driving protocol). Finally, the 
initial state of the units is arbitrary but uncorrelated. A sketch of the present setup is shown in 
Fig.~\ref{fig setup}. We remark that various extensions are possible as discussed at the end of this Letter. 

Below we will need the notion of the `Hamiltonian of mean force', an old concept~\cite{KirkwoodJCP1935} 
(see also Refs.~\cite{SeifertPRL2016, StrasbergEspositoPRE2019, CampisiTalknerHaenggiPRL2009}), which is defined 
via the reduced equilibrium state of a bipartite system $XB$. Specifically, 
\begin{equation}
 \pi_X^* \equiv \mbox{tr}_B\{\pi_{XB}\} \equiv \frac{e^{-\beta H_X^*}}{\C Z_X^*}, ~~~ 
 \C Z_X^* \equiv \frac{\C Z_{XB}}{\C Z_B}.
\end{equation}
Note that $\pi_X^*\neq\pi_X$ in general. In addition, $H_X^*$ depends on the inverse temperature $\beta$ and (possibly) 
a control parameter. Notice that the Hamiltonian of mean force for the system and all units simplifies as at any given 
time at most one unit is physically coupled to the system, e.g., for $t\in  I_k$, 
$H_{SU(\bb n)}^*(\lambda_t) = H_{SU(k)}^*(\lambda_t)$. Here and in general we use $U(\bb n)$ to denote the entire 
sequence of units from $U(0)$ to $U(n)$. 

The average rate of injected work (the power) has two contributions. For $t\in I_k$ we define 
\begin{align}
 \dot W_S(t)    &=  \lr{\frac{\partial H_S(\lambda_t)}{\partial t}}(t), \label{eq work normal}   \\
 \dot W_{SU(k)}(t) &=  \lr{\frac{\partial V_{SU(k)}(\lambda_t)}{\partial t}}(t), 
 \label{eq work SU}
\end{align}
where $\lr{\dots}(t)$ denotes a quantum statistical average at time $t$. It follows that the 
total mechanical work performed on the system up to time $t$ is 
\begin{align}
 W(t)  &= \int_{t_0^-}^t ds [\dot W_S(s) + \sum_k \dot W_{SU(k)}(s)]   \\
       &= \lr{H_\text{tot}(\lambda_t)}(t) - \lr{H_\text{tot}(\lambda_0^-)}(t_0^-).   \nonumber
\end{align}
Note that this definition of average mechanical work is widely accepted even in the strong coupling 
regime~\cite{StrasbergEspositoPRE2019, EspositoLindenbergVandenBroeckNJP2010, TakaraHasegawaDriebePLA2010, 
EspositoOchoaGalperinPRL2015, EspositoOchoaGalperinPRB2015, StrasbergEtAlNJP2016, LudovicoEtAlEntropy2016, 
StrasbergEtAlPRB2018, PerarnauLlobetEtAlPRL2018, WhitneyPRB2018, DouEtAlPRB2018} as it is directly related to the 
change in internal energy of the universe (i.e., the system, the bath and all units all together). 

\emph{Strong coupling repeated interactions framework.---} 
We start by introducing the basic concept of a nonequilibrium free energy adapted to the strong coupling 
regime~\cite{SeifertPRL2016, StrasbergEspositoPRE2019}, 
\begin{equation}\label{eq F}
 F_X(t) \equiv \mbox{tr}_X\{\rho_X(t)[H_X^*(\lambda_t) + \beta^{-1}\ln\rho_X(t)]\}.
\end{equation}
In the weak coupling limit, where $H_X^*(\lambda_t) \approx H_X(\lambda_t)$, this definition reduces to the 
conventional one. The slight modification allows us to express the second law even at strong 
coupling and even in presence of the system-unit interactions in the conventional way ($k_B\equiv1$): 
\begin{equation}\label{eq 2nd law RI}
 \Sigma(t) \equiv \beta[W(t) - \Delta F_{SU(\bb n)}(t)] \ge 0.
\end{equation}
Here, $\Sigma(t)$ denotes the entropy production and 
$\Delta F_{SU(\bb n)}(t) \equiv F_{SU(\bb n)}(t) - F_{SU(\bb n)}(t_0^-)$. Positivity of the second law 
follows by confirming that 
\begin{equation}\label{eq 2nd law RI 2}
 \Sigma(t) = D[\rho_\text{tot}(t)\|\pi_\text{tot}(\lambda_t)] - D[\rho_{SU(\bb n)}(t)\|\pi_{SU(\bb n)}^*(\lambda_t)],
\end{equation}
where $D[\rho\|\sigma] = \mbox{tr}\{\rho(\ln\rho-\ln\sigma)\} \ge 0$ is the quantum relative entropy. Hence, $\Sigma(t)$ 
is positive by monotonicity of relative entropy~\cite{UhlmannCMP1977, OhyaPetzBook1993}. The derivation uses only 
Eq.~(\ref{eq initial state}) and the unitary dynamics, which implies for the von Neumann entropy 
$S[\rho_\text{tot}(0)]\equiv-\mbox{tr}\{\rho_\text{tot}(0)\ln\rho_\text{tot}(0)\} = S[\rho_\text{tot}(t)]$. It is 
lengthy but straightforward and hence not displayed here. 

Equation~(\ref{eq 2nd law RI}) corresponds to the second law if we regard the system and \emph{all} units as one big 
system and explicitly keep their correlations in the description. In practice it often turns out that keeping the 
information about all units and all their correlations is superfluous (compare also with the discussion in 
Ref.~\cite{StrasbergEtAlPRX2017}). Thus, at time $t_{n+1}^-$ after the $n$'th interaction but before the $(n+1)$'th 
interaction, where the system is decoupled from all units, the following second law is practically more meaningful, 
\begin{align}
 \Sigma_S(t_{n+1}^-) 
 &= \beta[W(t_{n+1}^-) - \Delta F_S(t_{n+1}^-)] + \sum_{k=0}^{n}\Delta S[\rho_{U(k)}] \nonumber \\
  &\ge \Sigma(t_{n+1}^-) \ge 0. \label{eq 2nd law RI marginal} 
\end{align}
Here, we added a subscript $S$ to indicate that this is the entropy production from the system point of view 
ignoring superfluous information about the units. To arrive at Eq.~(\ref{eq 2nd law RI marginal}), we used 
subadditivity of entropy and $H_{U(k)} \sim 1_{U(k)}$. In contrast to Eq.~(\ref{eq 2nd law RI}) it contains only the 
change in the marginal von Neumann entropy of the units. The entropy production per interaction interval $I_n$ is then 
given by 
\begin{equation}
 \begin{split}\label{eq ent prod rate}
  \Sigma_S&(t_{n+1}^-) - \Sigma_S(t_n^-) = \\
  & \beta[W(t_{n+1}^-) - W(t_n^-) - F_S(t_{n+1}^-) + F_S(t_n^-)] \\
  & +S[\rho_{U(n)}(t_{n+1}^-)] - S[\rho_{U(n)}(t_n^-)].
 \end{split}
\end{equation}
It is the strong coupling generalization of the second law in the repeated interaction framework, see Eq.~(49) in 
Ref.~\cite{StrasbergEtAlPRX2017}. Interestingly, in contrast to the Markovian weak coupling situation, we cannot ensure 
the positivity of this expression. This is similar to the classical case~\cite{StrasbergEspositoPRE2019} and we will 
connect it to the notion of non-Markovianity later on. But first we will advance conceptually by introducing explicit 
measurements in the description. 

\emph{Quantum stochastic thermodynamics at strong coupling.---} 
We consider the case where the experimenter measures the state of the unit \emph{after} the interaction with the system 
as indicated in Fig.~\ref{fig setup} (i.e., $V_{SU(n)}(\lambda_t) = 0$ at the time $t$ of the measurement). By doing so, 
she can gather valuable information about the state of the system. In a moment we will also show that this allows her to 
implement arbitrary generalized measurements on the system and that the resulting theory can be fruitfully linked to the 
study of quantum stochastic processes and quantum causal models. 

Mathematically, we denote the measurement result of the $k$'th unit by $r_k$ and associate a positive operator 
$P_{r_k}$ to it, which fulfills the normalization condition $\sum_{r_k} P_{r_k}^2 = 1_{U(k)}$. The state of the unit then 
changes according to the map $\C P_{r_k}\rho_{U(k)} \equiv P_{r_k}\rho_{U(k)} P_{r_k} \equiv \tilde\rho_{U(k)}(r_k)$. 
Notice that $\tilde\rho_{U(k)}(r_k)$ is a subnormalized state with the probability 
$p(r_k) = \mbox{tr}_{U(k)}\{\tilde\rho_{U(k)}(r_k)\}$ as its norm. After multiple units were subjected to their 
respective measurements giving results $\bb r_n \equiv (r_n,\dots,r_1,r_0)$, the global subnormalized state reads 
\begin{equation}
 \tilde\rho_\text{tot}(\bb r_n,t_{n+1}^-) 
 = \left(\prod_{k=0}^n \C P_{r_k}\C U_{k+1,k}\right) \rho_\text{tot}(t_0^-).
\end{equation}
Here, $\C U_{k+1,k}$ describes the global unitary evolution from $t_k^-$ (shortly before the $k$'th unit starts 
interacting with the system) to $t_{k+1}^-$. Due to the fact that the measurement always acts after the 
interaction we can also write 
\begin{equation}\label{eq dynamics measured}
 \tilde\rho_\text{tot}(\bb r_n,t_{n+1}^-) 
 = \left(\prod_{k=0}^n \C P_{r_k}\right) \rho_\text{tot}(t_{n+1}^-),
\end{equation}
where $\rho_\text{tot}(t_{n+1}^-) = \C U_{n+1,0} \rho_\text{tot}(t_0^-)$ is the global time-evolved state without 
any measurements. This allows us to confirm the useful relation
\begin{equation}\label{eq average equals unmeasured}
 \sum_{\bb r_n} \tilde\rho_{SB}(\bb r_n,t_{n+1}^-) = \rho_{SB}(t_{n+1}^-),
\end{equation}
i.e., the average system-bath state does not change due to the measurements. This is not true for the units. 

Inspired by Ref.~\cite{StrasbergPRE2019}, we now introduce the following thermodynamic definitions along a single 
trajectory characterized by the measurement results $\bb r_n$. First, for $t\in I_n$ the stochastic power 
$\dot w_S(\bb r_n,t)$ and $\dot w_{SU(n)}(\bb r_{n-1},t)$ are simply obtained from Eqs.~(\ref{eq work normal}) 
and~(\ref{eq work SU}) by replacing the average over the unconditional state $\rho_{SU(\bb n)}(t)$ with an average over 
the conditional state $\rho_{SU(\bb n)}(\bb r_n,t)$. Notice that $\dot w_{SU(n)}(\bb r_{n-1},t)$ does not depend on the 
last measurement outcome $r_n$ because by construction the measurement $\C P_{r_n}$ acts after the $n$th unit has 
interacted with the system. Therefore, together with Eq.~(\ref{eq average equals unmeasured}) we immediately obtain the 
relations $\sum_{\bb r_n} p(\bb r_n) \dot w_S(\bb r_n,t) = \dot W_S(t)$ and 
$\sum_{\bb r_n} p(\bb r_n) \dot w_{SU(n)}(\bb r_{n-1},t) = \dot W_{SU(n)}(t)$. Second, we have to generalize the 
nonequilibrium free energy to the stochastic case, which becomes 
\begin{equation}
 \begin{split}
  f_{SU(\bb n)}(\bb r_n,t) 
  \equiv&~ \lr{H_{SU(\bb n)}^*(\lambda_t)}(\bb r_n,t) \\
  & + T\left\{\ln p(\bb r_n) - S[\rho_{SU(\bb n)}(\bb r_n,t)]\right\},
 \end{split}
\end{equation}
where $\lr{\dots}(\bb r_n,t)$ denotes an average with respect to the conditional state. An essential difference 
compared to definition~(\ref{eq F}) is the appearence of the stochastic entropy $-\ln p(\bb r_n)$ associated to the 
measurement results obtained with probability 
$p(\bb r_n) = \mbox{tr}\{\tilde\rho_\text{tot}(\bb r_n,t)\}$. A similar but not identical 
construction is used in classical stochastic thermodynamics~\cite{SeifertPRL2005}, compare with the discussion in 
Refs.~\cite{StrasbergPRE2019, StrasbergWinterPRE2019}. In contrast to the stochastic work, we have in general 
$\sum_{\bb r_n} p(\bb r_n) f_{SU(\bb n)}(\bb r_n,t) \neq F_{SU(\bb n)}(t)$. Finally, we introduce the stochastic 
entropy production 
\begin{equation}
 \sigma(\bb r_n,t) \equiv \beta[w(\bb r_n,t) - \Delta f_{SU(\bb n)}(\bb r_n,t)].
\end{equation}
As in classical stochastic thermodynamics, it can be negative along a single trajectory~\cite{SeifertRPP2012, 
VandenBroeckEspositoPhysA2015}. However, we will now prove that on average 
$\sum_{\bb r_n} p(\bb r_n) \sigma(\bb r_n,t) \ge 0$,
which demonstrates the thermodynamic consistency of our strong coupling quantum stochastic framework. 

As a consequence of Eq.~(\ref{eq average equals unmeasured}) and our previously derived second 
law~(\ref{eq 2nd law RI}), we confirm that 
\begin{align}
 & \sum_{\bb r_n} p(\bb r_n) \sigma(\bb r_n,t) - \Sigma(t) = \label{eq diff 2nd laws} \\
 & \sum_{\bb r_n} p(\bb r_n) \left\{S[\rho_{SU(\bb n)}(\bb r_n,t)] - \ln p(\bb r_n)\right\} - S[\rho_{SU(\bb n)}(t)]. \nonumber
\end{align}
This quantifies the change in informational entropy of all constituents (system, units and the classical memory) 
due to the big joint measurement~(\ref{eq dynamics measured}). Its positivity follows from the Lemma in 
Ref.~\cite{StrasbergPRE2019}, which simply combines Theorem~11 of Ref.~\cite{JacobsBook2014} and Theorem~11.10 of 
Ref.~\cite{NielsenChuangBook2000} and which can be interpreted as the second law for a quantum measurement. 
Hence, we conclude 
\begin{equation}\label{eq 2nd law traj cor}
 \sum_{\bb r_n} p(\bb r_n) \sigma(\bb r_n,t) \ge \Sigma(t) \ge 0.
\end{equation}
As before, the (averaged) stochastic entropy production $\sigma(\bb r_n,t)$ contains the information about 
all the correlations in the units, which is typically not needed. Using subadditivity of entropy, it is again 
possible to arrive at expressions similar to Eq.~(\ref{eq 2nd law RI marginal}). Note that, depending on the 
experimental situation, one could decide to not only discard information about the unit correlations, but also 
about the measurement results $\bb r_n$. 

This concludes the formal part of the Letter, where we have introduced a consistent notion of work, nonequilibrium free 
energy and entropy production along a single run of an experiment regardless of any details of the system-bath 
coupling. It is instructive to connect the present picture to the theory of quantum causal models and quantum 
stochastic processes. If we consider the limit of an instantaneous system-unit interaction, ideally described 
by a coupling of the form $V_{SU(k)}(\lambda_t) = v_k\delta(t-t_k)$, we can write the system-bath dynamics as 
\begin{equation}\label{eq dynamics SB trajectory}
 \tilde\rho_{SB}(\bb r_n,t_{n+1}^-) = \left(\prod_{k=0}^n \C U^{SB}_{k+1,k}\C A_{r_k}\right) \rho_{SB}(t_0^-).
\end{equation}
Here, $\C U^{SB}_{k+1,k}$ is the unitary time evolution generated by $H_{SB}(\lambda_t)$ and the completely positive 
map $\C A_{r_k}$ is defined via its action 
$\C A_{r_k}\rho_S = \mbox{tr}_U\{\C P_{r_k}[e^{-iv_k\hbar}\rho_S\rho_{U(k)}e^{iv_k/\hbar}]\}$ on an 
arbitrary system state $\rho_S$. In this context $\C A_{r_k}$ is also known as an `instrument' describing the most 
general state transformation possible in quantum mechanics~\cite{HolevoBook2001b, WisemanMilburnBook2010}. The 
application of a set of instruments $\C A_{r_0},\dots,\C A_{r_n}$ to an open quantum system \emph{defines} a general 
quantum stochastic process (or quantum causal model), which can be formally represented by a `quantum comb' or `process 
tensor'~\cite{ChiribellaDArianoPerinottiPRL2008, ChiribellaDArianoPerinottiPRA2009, CostaShrapnelNJP2016, 
OreshkovGiarmatziNJP2016, PollockEtAlPRA2018, MilzEtAlArXiv2017}. The sole difference compared to the most general 
case is that we do \emph{not} allow for real-time feedback control, i.e., the instruments $\C A_{r_k}$ are not allowed 
to depend on the previous results $\bb r_{k-1}$, otherwise Eq.~(\ref{eq average equals unmeasured}) would no longer be 
true. Whether the present framework can be extended to arbitrary real-time feedback control as in the Markovian 
case~\cite{StrasbergPRE2019, StrasbergWinterPRE2019} remains an open question. 

\emph{Thermodynamic signatures of non-Markovianity.---} 
We now turn towards an important application linking the field of quantum thermodynamics and quantum 
non-Markovianity~\cite{RivasHuelgaPlenioRPP2014, BreuerEtAlRMP2016} in a rigorous way. As recognized below 
Eq.~(\ref{eq ent prod rate}), at strong coupling we cannot ensure that the entropy production is positive 
in every time-interval $[t_\ell,t_k)$ for $t_\ell>t_k>t_0$. There have been repeated claims in the literature that 
negative entropy production rates indicate non-Markovianity~\cite{ArgentieriEtAlEPL2014, BhattacharyaEtAlPRA1017, 
MarcantoniEtAlSR2017, PopovicVacchiniCampbellPRA2018, ThomasEtAlPRE2018, BhattacharyaBhattacharyaMajumdararXiv2019}.
Doubts were raised in Refs.~\cite{StrasbergEspositoPRE2019} since the definitions for entropy production rates used 
in Refs.~\cite{ArgentieriEtAlEPL2014, BhattacharyaEtAlPRA1017, MarcantoniEtAlSR2017, PopovicVacchiniCampbellPRA2018, 
ThomasEtAlPRE2018, BhattacharyaBhattacharyaMajumdararXiv2019} do not yield an overall positive entropy production 
when integrated from the initial time $t_0$ to any final time $t>t_0$. Moreover, they can be even negative for 
Markovian dynamics~\cite{StrasbergEspositoPRE2019}. 

On the other hand, for a suitable notion of entropy production based on the Hamiltonian of mean force progress was 
achieved for classical dynamics~\cite{StrasbergEspositoPRE2019}. In there, the situation of a strongly coupled 
system prepared in an arbitrary nonequilibrium state was considered (no repeated interactions were present). Then, it 
was shown that Markovian dynamics necessarily imply a positive entropy production rate if the system is \emph{undriven} 
(i.e., $\lambda_t = $ constant). In our quantum formalism the system is initially in equilibrium such that its state 
does not change when undriven and left on its own. However, we can link the present picture to the classical case 
by realizing that we can use the very first unit $U(0)$ to prepare the system in an arbitrary nonequilibrium state via a 
short control operation. This preparation procedure has a thermodynamic cost captured by the always positive entropy 
production $\Sigma_S(t) = \beta W_{SU}(t) - \beta\Delta F_S(t) - \beta \Delta S[\rho_{U(0)}] \ge 0$, compare with 
Eq.~(\ref{eq 2nd law RI marginal}). After the system-unit interaction, the system is left on its own and the entropy 
production in between any two times $t_2>t_1>t_0$ reads 
\begin{equation}
 \begin{split}\label{eq EP rate}
  \Sigma_S(t_2) - \Sigma_S(t_1) &=  -\beta[F_S(t_2)-F_S(t_1)]   \\
  &= D[\rho_S(t_1)\|\pi_S^*] - D[\rho_S(t_2)\|\pi_S^*].
 \end{split}
\end{equation}
This quantifies the dissipation associated with the relaxation dynamics of the system. Equation~(\ref{eq EP rate}) is 
positive if the dynamics are Markovian \emph{and} if $\pi_S^*$ is a steady state of the dynamics at any time $t$. 
Interestingly, the latter point can be shown rigorously based on the definition of Markovianity from 
Ref.~\cite{PollockEtAlPRL2018}, which is adapted to the situation of a general quantum stochastic process as used here. 
This is proven in the Appendix~\ref{sec steady state}. Thus, $\Sigma_S(t_2) - \Sigma_S(t_1) \ge 0$ for a Markov process 
in complete analogy to the classical result~\cite{StrasbergEspositoPRE2019}. This opens up the door to investigate the 
interplay between entropy production and non-Markovianity in a mathematically and thermodynamically rigorous sense for 
quantum systems. 

\emph{Further applications.---} 
The ability to analyse general non-Markovian quantum processes from a thermodynamic perspective will find applications 
in various areas. One example is sequential quantum metrology~\cite{GiovannetiLloydMacconeNatPho2011}. Specifically, 
a particular intriguing parameter to estimate is the temperature of a system. Much progress has been achieved to 
understand it from the perspective of metrology~\cite{MehboudiSanperaCorreaJPA2019}, but the thermodynamic costs of 
thermometry were not yet explored. With the recent progress in the design of optimal quantum 
probes~\cite{CorreAtAlPRL2015} and strong-coupling thermometry~\cite{CorreaEtAlPRA2017}, the present Letter opens up 
the possibility to thermodynamically analyse many scenarios in metrology and thermometry. Furthermore, recent progress 
shows how to unambigously detect quantum features in quantum stochastic processes~\cite{SmirneEtAlQST2018, 
StrasbergDiazPRA2019, MilzEgloffEtAlArXiv2019}. Since quantum thermodynamics is still in the search for clear 
observable quantum effects induced by coherence~\cite{OnamGonzalezEtAlPRE2019}, the present letter will allow to 
rigorously address such questions. Furthermore, originally used to understand Nobelprize-winning 
experiments~\cite{SayrinEtAlNature2011, ZhouEtAlPRL2012} in quantum optics from a thermodynamic 
perspective~\cite{StrasbergPRE2019}, the present framework can be used to explore more general cases where the system 
is not a high-quality cavity and has substantial losses or is coupled to other cavities. Also the units do not have to 
be identical, which opens up the possibility to, e.g., thermodynamically analyse single-photon distillation 
experiments~\cite{DaissEtAlPRL2019} where an atom in a cavity is first probed by a weak coherent pulse (unit 1) followed 
by a measurement (modeled by unit 2) to herald the photon distillation. Quite generally, the present setup is even 
relevant for experiments in the Markovian regime, if detailed control about all system parts is not possible. Finally, 
the present framework can be combined with the traditional picture of scattering theory and, following 
Ref.~\cite{StrasbergEtAlPRX2017}, it allows to investigate Maxwell's demon and Landauer's principle at strong coupling.

\emph{Extensions.---} 
As detailed in Appendix~\ref{sec extensions}, the present framework can be extended into various directions: the 
Hamiltonian of the units does not need to be degenerate, the initial state of the units can be correlated, and the 
general identities~(\ref{eq 2nd law RI}),~(\ref{eq 2nd law RI marginal}) and~(\ref{eq 2nd law traj cor}) still hold in 
the case where the system-bath coupling is time-dependent. In the last case, however, the theory is no longer 
`operational' in the sense that explicit knowledge about the state of the bath is necessary, which is hardly accessible. 
Similarly, we show in Appendix~\ref{sec extensions} how to treat multiple heat baths. Unfortunately, also in that case 
the state of the system and units does not suffice to have access to all thermodynamic quantities~\footnote{This 
problem, however, is already present in classical stochastic thermodynamics~\cite{SeifertRPP2012, 
VandenBroeckEspositoPhysA2015}, see also the discussion in Ref.~\cite{StrasbergPRE2019}.}. 

\emph{Concluding remarks.---} 
The present contribution establishes a consistent thermodynamic framework -- even along a single trajectory recorded in 
an experiment -- for a system in contact with an arbitrary bath and additionally subjected to arbitrary nonequilibrium 
resources interacting one by one with the system. This pushes the applicability of nonequilibrium thermodynamics far 
beyond its traditional scope. Furthermore, the present work also demonstrates how quantum stochastic thermodynamics 
departs from its classical version in 
the strong coupling regime~\cite{SeifertPRL2016, JarzynskiPRX2017, MillerAndersPRE2017, StrasbergEspositoPRE2017, 
StrasbergEspositoPRE2019}. While the basic concepts at the unmeasured level are similar, any possible measurement 
strategy has a non-trivial influence on the description in the quantum regime, even on average. For instance, in 
general there is a strict inequality on the left hand side of Eq.~(\ref{eq 2nd law traj cor}). This is not a deficiency 
of our theory, but a \emph{necessary} ingredient, which can be already recognized at the level of the work 
statistics~\cite{PerarnauLlobetEtAlPRL2017}. Quantum stochastic thermodynamics is \emph{more} than a mere extension of 
its classical counterpart. The present operational approach is, however, flexible enough to reproduce the unmeasured 
picture: it is recovered by choosing the trivial but legitimiate measurement operator $P_{r_k} = 1_{U(k)}$, i.e., the 
identity. Then, the stochastic entropy production $\sigma(\bb r_n)$ reduces to $\Sigma$. 

\emph{Acknowledgments.---} 
I thank Kavan Modi for illuminating initial discussions. This research was financially supported by the DFG (project 
STR 1505/2-1) and also the Spanish MINECO FIS2016-80681-P (AEI-FEDER, UE). 


\bibliography{/home/philipp/Documents/references/books,/home/philipp/Documents/references/open_systems,/home/philipp/Documents/references/thermo,/home/philipp/Documents/references/info_thermo,/home/philipp/Documents/references/general_QM,/home/philipp/Documents/references/math_phys,/home/philipp/Documents/references/equilibration}

\appendix
\begin{widetext}
\section{Steady state of an undriven quantum Markov process}
\label{sec steady state}

According to Ref.~\cite{PollockEtAlPRL2018}, a Markov process is characterized by a set of completely positive and 
trace-preserving maps $\{\Lambda(t_\ell,t_k)|t_\ell>t_k\}$ such that the normalized state conditioned on an arbitrary 
sequence of control operations $\C A_0,\C A_1,\dots,\C A_n$ can be written as 
\begin{equation}
 \rho(t_n^+|\C A_n,\dots,\C A_1,\C A_0) = \C A_n\Lambda(t_n,t_{n-1})\dots\C A_1\Lambda(t_1,t_0)\C A_0\rho_S(t_0^-).
\end{equation}
Notice that the composition law $\Lambda(t_3,t_1) = \Lambda(t_3,t_2) \Lambda(t_2,t_1)$ for $t_3>t_2>t_1$ 
(the `quantum Chapman-Kolmogorov equation') automatically follows from that by realizing that the identity operation 
$\C I$ is a legitimate control operation too.

By applying this to our problem, we immediately realize that we obtain the relation 
\begin{equation}
 \Lambda(t_k,t_0)\pi_S^* = \pi_S^*
\end{equation}
for any $t_k > t_0$ and hence, the same holds for any intermediate map $\Lambda(t_\ell,t_k)$ ($t_\ell>t_k$). 
Note that the assumption of no driving as well as the particular form of the initial system-bath state 
[$\rho_{SB}(t_0^-) = \pi_{SB}$] is crucial to arrive at this conclusion. Equation~(20) of the main text can therefore 
be written as 
\begin{equation}
 D[\rho_S(t_1)\|\pi_S^*] - D[\rho_S(t_2)\|\pi_S^*] = D[\rho_S(t_1)\|\pi_S^*] - D[\Lambda(t_2,t_1)\rho_S(t_1)\|\Lambda(t_2,t_1)\pi_S^*].
\end{equation}
Now, we can use that for a Markov process relative entropy is contractive~\cite{RivasHuelgaPlenioRPP2014, 
BreuerEtAlRMP2016} to conclude that Eq.~(20) of the main text is positive. 

It is, of course, worth to point out that in general, for an initially correlated state of the form $\pi_{SB}$, 
there exists \emph{no} such set of completely positive and trace-preserving maps $\{\Lambda(t_\ell,t_k)|t_\ell>t_k\}$. 
Obviously, this simply shows that one should not typically expect Markovian 
behaviour in strongly coupled systems. Nevertheless, in particular -- and not necessarily uninteresting -- limiting 
cases this can still be the case, e.g., in the limit of time-scale separation~\cite{StrasbergEspositoPRE2017}, 
which allows in the quantum case to derive a Markovian master equation in a Polaron frame~\cite{SchallerEtAlNJP2013, 
GelbwaserKlimovskyAspuruGuzikJPCL2015, StrasbergEtAlNJP2016}. 

\section{Extensions of strong coupling quantum stochastic thermodynamics}
\label{sec extensions}

For later comparison, we start by listing the definitions of internal energy, heat flow and system entropy complementary 
to the definition of work and nonequilibrium free energy based on the assumptions of the main text (only one heat 
bath, energy degenerate units, no driving in the system-bath interaction). At the unmeasured level we have 
\begin{align}
 E^*_{SU(\bb n)}(t) &\equiv \lr{H_{SU(\bb n)}^*(\lambda_t) + \beta\partial_\beta H_{SU(\bb n)}^*(\lambda_t)}(t), \label{eq def int energy basic} \\
 Q_{SU(\bb n)}(t) &\equiv \Delta E^*_{SU(\bb n)}(t) - W(t), \\
 S_{SU(\bb n)}(t) &\equiv S[\rho_{SU(\bb n)}(t)] + \beta^2\lr{\partial_\beta H_{SU(\bb n)}^*(\lambda_t)}(t).
\end{align}
Remember that $W(t)$ is the total work as defined in Eq.~(6) of the main text and the letter $S$ without subscripts 
always denotes the von Neumann entropy $S(\rho) = -\mbox{tr}\{\rho\ln\rho\}$. Hence, we see that the thermodynamic 
entropy $S_{SU(\bb n)}(t)$ is not identical to the von-Neumann entropy in the strong coupling regime. Also note that 
$F_{SU(\bb n)}(t) = E^*_{SU(\bb n)}(t) - TS_{SU(\bb n)}(t)$ reproduces the 
definition of nonequilibrium free energy of the main text. The above definitions have been already discussed for a 
single system strongly coupled to a bath in the classical~\cite{SeifertPRL2016, JarzynskiPRX2017, MillerAndersPRE2017, 
StrasbergEspositoPRE2017, StrasbergEspositoPRE2019} and quantum~\cite{StrasbergEspositoPRE2019} case in the absence of 
any units. The second law of nonequilibrium thermodynamics can be expressed in the two alternative forms 
\begin{equation}\label{eq 2nd law basic}
 \Sigma(t) = \beta[W(t) - \Delta F_{SU(\bb n)}(t)] 
 = \Delta S_{SU(\bb n)}(t) - \beta Q_{SU(\bb n)}(t) \ge 0.
\end{equation}

To extend the above definitions to the trajectory level, we have to replace the state $\rho_{SU(\bb n)}(t)$ by the 
conditional state $\rho_{SU(\bb n)}(\bb r_n,t)$ and we have to add the stochastic entropy of the detector 
$[-\ln p(\bb r_n)]$ to the definition of entropy. We denote this as 
\begin{align}
 e^*_{SU(\bb n)}(\bb r_n,t) &\equiv \lr{H_{SU(\bb n)}^*(\lambda_t) + \beta\partial_\beta H_{SU(\bb n)}^*(\lambda_t)}(\bb r_n,t), \\
 q_{SU(\bb n)}(\bb r_n,t) &\equiv \Delta e^*_{SU(\bb n)}(\bb r_n,t) - w(\bb r_n,t), \label{eq heat stochastic basic} \\
 s_{SU(\bb n)}(\bb r_n,t) &\equiv S[\rho_{SU(\bb n)}(\bb r_n,t)] + \beta^2\lr{\partial_\beta H_{SU(\bb n)}^*(\lambda_t)}(\bb r_n,t) - \ln p(\bb r_n).
\end{align}
Note that we naturally assume the notation $(\bb r_n,t)$ to imply a time $t$ where we have already measured the state 
of the $n$th unit (giving result $r_n$) but where the next $(n+1)$th unit was not yet measured. 
The stochastic entropy production can be again expressed in the two alternative forms 
\begin{equation}
 \sigma(\bb r_n,t) = \beta[w(\bb r_n,t) - \Delta f_{SU(\bb n)}(\bb r_n,t)] 
 = \Delta s_{SU(\bb n)}(\bb r_n,t) - \beta q_{SU(\bb n)}(\bb r_n,t).
\end{equation}
On average, $\sum_{\bb r_n} p(\bb r_n)\sigma(\bb r_n,t) \ge 0$. 

We will now investigate various extensions of the framework presented in the main text. We will be very explicit with 
the steps in Extension~4 (multiple heat baths) as it is the most general setting implying all the others 
(including the one of the main text). Therefore, for the Extensions~1 to~3 we limit ourselves to only pointing out some 
essential observations. 

\subsection{Extension 1: Driven system-bath interaction}
\label{sec ext 1}

In this case we allow that the system-bath interaction $V_{SB} = V_{SB}(\lambda_t)$ depends on some externally 
prescribed time-dependent protocol $\lambda_t$. Such a setting is important, e.g., to model how a system is put into 
contact with a bath. Including this case only changes the definition of power [Eq. (4) in the main text] in an obvious 
way to 
\begin{equation}\label{eq work with SB}
 \dot W_S(t) = \mbox{tr}_{SB}\left\{\left[\frac{\partial H_S(\lambda_t)}{\partial t} + \frac{\partial V_{SB}(\lambda_t)}{\partial t}\right]\rho_{SB}(t)\right\}.
\end{equation}
Despite being only a little formal change, this scenario implies that the power can no longer be computed based only 
on the knowledge of the system state $\rho_S(t)$, but in general requires to know the entire state $\rho_{SB}(t)$. 
This is problematic from an experimental point of view as the theory is no longer operational and also from a 
computational point of view it is very challenging. The stochastic generalization $\dot w_S(\bb r_n,t)$ of 
Eq.~(\ref{eq work with SB}) simply uses the conditional state $\rho_{SB}(\bb r_n,t)$. 

\subsection{Extension 2: Correlated initial unit state}

As will become clear in the detailed derivation below, the initial state of the units $\rho_{U(\bb n)}(t_0^-)$ can 
indeed be arbitrary and does not need to be decorrelated. However, in practical applications this is often unnecessary 
and only complicates the theoretical treatment. Furthermore, deriving a simplified second law for the system alone 
[as in Eqs.~(10) and~(11) of the main text] becomes non-trivial then. 

\subsection{Extension 3: Energetic units}

This generalization is slightly more subtle as the previous two when it comes to the trajectory level. The reason 
is that for energetically non-degenerate units the act of measurement, which requires to couple the microscopic unit to 
some macroscopic (not further specified) detection apparatus, can change the energy of the unit even on average unless 
the measurement operators $P_{r_k}$ commute with the unit Hamiltonian or the final unit state. While this is often 
the case in practice, see, e.g., Refs.~\cite{SayrinEtAlNature2011, ZhouEtAlPRL2012}, we are here interested in the most 
general scenario. Note that these subtleties were already discussed in greater detail in Ref.~\cite{StrasbergPRE2019}, 
compare also with the related (but not identical) concept of ``quantum heat'' introduced in 
Ref.~\cite{ElouardEtAlQInf2017}. 

To start with, nothing changes at the unmeasured level: Eqs.~(\ref{eq def int energy basic}) to~(\ref{eq 2nd law basic}) 
remain true as usual. Now, to isolate the contribution steming solely from the energetic measurements of the unit 
due to the final measurement we define 
\begin{equation}\label{eq meas heat}
 q_\text{meas}(\bb r_n,t) \equiv e_{U(\bb n)}(\bb r_n,t) - E_{U(\bb n)}(t).
\end{equation}
Here, $E_{U(\bb n)}(t) = \mbox{tr}_{U(\bb n)}\{H_{U(\bb n)}\rho_{U(\bb n)}(t)\}$ is the energy of all units 
evaluated with respect to the unmeasured state 
$\rho_{U(\bb n)}(t) = \mbox{tr}_{SB}\{\C U_{t,t_0}\rho_\text{tot}(t_0^-)\}$ of the units, whereas 
$e_{U(\bb n)}(t) = \mbox{tr}_{U(\bb n)}\{H_{U(\bb n)}\rho_{U(\bb n)}(\bb r_n,t)\}$ involves the measured state, 
compare also with Eq.~(13) in the main text. Note that at the time of the measurement of the $n$th unit giving result 
$r_n$, the unit is physically decoupled from the system, i.e., we have $V_{SU(n)}(\lambda_t) = 0$ as indicated in 
Fig.~1 of the main text. We also remark that it is debated whether Eq.~(\ref{eq meas heat}) should be counted as `heat' 
or `work'. In fact, we will now split off this term from the total stochastic heat~(\ref{eq heat stochastic basic}) 
as follows 
\begin{equation}
 q_{SU(\bb n)}(\bb r_n,t) \equiv q(\bb r_n,t) + q_\text{meas}(\bb r_n,t) 
 = \Delta e^*_{SU(\bb n)}(\bb r_n,t) - w(\bb r_n,t).
\end{equation}
This defines $q(\bb r_n,t)$ as the essential part of the stochastic heat flow, which contains two contributions: first, 
all the heat flow affecting the system \emph{and} unit during the time without measurements and second, any stochastic 
change of the system energy due to the measurement, which causes an update of our state of knowledge about the system. 
Consequently, using Eq.~(14) of the main text we can confirm that on average 
\begin{equation}
 \sum_{\bb r_n} p(\bb r_n)q(\bb r_n,t) = \Delta E^*_{SU(\bb n)}(t) - W(t) = Q_{SU(\bb n)}(t).
\end{equation}
We then define the stochastic entropy production by including only the heat flow $q$ and we exlude $q_\text{meas}$, 
i.e., 
\begin{equation}
 \sigma(\bb r_n,t) = \Delta s_{SU(\bb n)}(\bb r_n,t) - \beta q(\bb r_n,t).
\end{equation}
This definition allows us to conclude that $\sum_{\bb r_n} p(\bb r_n) \sigma(\bb r_n,t) \ge 0$, see the next Section for 
all the details. Note that in case of energetically degenerate units we always have $q_\text{meas}(\bb r_n,t) = 0$ and 
$q_{SU(\bb n)}(\bb r_n,t) = q(\bb r_n,t)$. 

\subsection{Extension 4: Multiple heat baths}

The setup we are here assuming is specified by the following Hamiltonian 
\begin{equation}
 H_\text{tot}(\lambda_t) = H_S(\lambda_t) + V_{SB_1}(\lambda_t) + H_{B_1} + V_{SB_2}(\lambda_t) + H_{B_2} 
 + \sum_{k=0}^n [V_{SU(k)}(\lambda_t) + H_{U(k)}]
\end{equation}
and initial state 
\begin{equation}\label{eq initial state general}
 \rho_\text{tot}(t_0^-) = \pi_{SB_1}(\beta_1,\lambda_t)\otimes\pi_{B_2}(\beta_2)\otimes\rho_{U(\bb n)}(t_0^-).
\end{equation}
Physically, it describes a system coupled to two heat baths $B_1$ and $B_2$, initialized at inverse temperatures 
$\beta_1$ and $\beta_2$ respectively, which additionally interacts with a stream of units as specified in the main text. 
The initial state is chosen such that the system is in a joint equilibrium state 
with $B_1$ decoupled from the equilibrium state of $B_2$. We thus assume that the coupling with $B_2$ is suddenly 
switched on at the initial time $t_0$. Note that decorrelated system-bath states are commonly assumed in the theory 
of open quantum systems. The initial state of the units is arbitrary but decorrelated from the system and the baths. 

We start with the thermodynamic description at the unmeasured level, which essentially combines the tools of 
Refs.~\cite{EspositoLindenbergVandenBroeckNJP2010, TakaraHasegawaDriebePLA2010, SeifertPRL2016, 
StrasbergEspositoPRE2019}. We therefore use the following basic definitions: 
\begin{align}
 E_{SU(\bb n)}^*(t) &= \lr{H_{SU(\bb n)}^*(\lambda_t) + \beta_1\partial_{\beta_1} H_{SU(\bb n)}^*(\lambda_t) + V_{SB_2}(\lambda_t)}(t), \\
 W(t) &= \int_{t_0^-}^t ds \lr{\frac{\partial H_\text{tot}(\lambda_s)}{\partial s}} 
 = \lr{H_\text{tot}(\lambda_t)}(t) - \lr{H_\text{tot}(\lambda_0^-)}(t_0^-), \\
 Q_1(t) &= \Delta E^*_{SU(\bb n)}(t) - W(t) - Q_2(t), \\
 Q_2(t) &= -[\lr{H_{B_2}}(t) - \lr{H_{B_2}}(t_0^-)], \\
 S_{SU(\bb n)}(t) &= S[\rho_{SU(\bb n)}(t)] + \beta_1^2\lr{\partial_{\beta_1} H_{SU(\bb n)}^*(\lambda_t)}(t).
\end{align}
Note that the internal energy differs from the previous definition~(\ref{eq def int energy basic}) by explicitly 
including the interaction $V_{SB_2}$ between the system and $B_2$~\cite{EspositoLindenbergVandenBroeckNJP2010}. 
Furthermore, the heat flow from $B_2$ is defined as (minus) the change in the expectation value of $H_{B_2}$. 
Finally, remember that the Hamiltonian of mean force depends explicitly on the inverse temperature $\beta_1$ of $B_1$, 
which we suppress in the notation. We will now show that the second law as known from phenomenological nonequilibrium 
thermodynamics holds: 
\begin{equation}
 \Sigma(t) = \Delta S_{SU(\bb n)}(t) - \beta_1Q_1(t) - \beta_2Q_2(t) \ge 0.
\end{equation}
For this purpose we notice that the heat flux from $B_1$ can be microscopically expressed as 
\begin{equation}
 Q_1(t) = \Delta \lr{H_{SU(\bb n)}^*(\lambda_t)}(t) + \beta_1\Delta\lr{\partial_{\beta_1} H_{SU(\bb n)}^*(\lambda_t)}(t) -\Delta\lr{H_{SB_1U(\bb n)}(\lambda_t)}(t),
\end{equation}
where the delta-notation means $\Delta f(\lambda_t,t) = f(\lambda_t,t) - f(\lambda_0^-,t_0^-)$ for any $f$ 
and where we defined $H_{SB_1U(\bb n)}(\lambda_t) \equiv H_{SU(\bb n)}(\lambda_t) + V_{SB_1}(\lambda_t) + H_{B_1}$. The 
entropy production therefore reads 
\begin{equation}
 \Sigma(t) = \Delta S[\rho_{SU(\bb n)}(t)] - \beta_1\Delta \lr{H_{SU(\bb n)}^*(\lambda_t)}(t) + \beta_1 \Delta\lr{H_{SB_1U(\bb n)}(\lambda_t)}(t) + \beta_2\Delta\lr{H_{B_2}}(t).
\end{equation}
Next, we use the standard trick $X = \ln e^X$ to write 
\begin{equation}
 \begin{split}
  \Sigma(t) =&~ 
  \Delta S[\rho_{SU(\bb n)}(t)] 
  + \mbox{tr}\left\{\rho_{SU(\bb n)}(t)\ln e^{-\beta_1 H_{SU(\bb n)}^*(\lambda_t)} 
     - \rho_{SU(\bb n)}(t_0^-)\ln e^{-\beta_1 H_{SU(\bb n)}^*(\lambda_0^-)}\right\} \\ 
  & - \mbox{tr}\left\{\rho_{SB_1U(\bb n)}(t)\ln e^{-\beta_1 H_{SB_1U(\bb n)}(\lambda_t)} 
     - \rho_{SB_1U(\bb n)}(t_0^-)\ln e^{-\beta_1 H_{SB_1U(\bb n)}(\lambda_0^-)}\right\} \\
  & - \mbox{tr}\left\{[\rho_{B_2}(t) - \rho_{B_2}(t_0^-)]\ln e^{-\beta_2 H_{B_2}}\right\} \\
  =&~ \Delta S[\rho_{SU(\bb n)}(t)] 
  + \mbox{tr}\left\{\rho_{SU(\bb n)}(t)\ln\pi^*_{SU(\bb n)}(\beta_1,\lambda_t) 
     - \rho_{SU(\bb n)}(t_0^-)\ln\pi^*_{SU(\bb n)}(\beta_1,\lambda_0^-)\right\} \\ 
  & - \mbox{tr}\left\{\rho_{SB_1U(\bb n)}(t)\ln\pi_{SB_1U(\bb n)}(\beta_1,\lambda_t) 
     - \rho_{SB_1U(\bb n)}(t_0^-)\ln\pi_{SB_1U(\bb n)}(\beta_1,\lambda_0^-)\right\} \\
  & - \mbox{tr}\left\{[\rho_{B_2}(t) - \rho_{B_2}(t_0^-)]\ln\pi_{B_2}(\beta_2)\right\} 
     + \ln\frac{\C Z^*_{SU(\bb n)}(\beta_1,\lambda_t)\C Z_{SB_1U(\bb n)}(\beta_1,\lambda_0^-)\C Z_{B_2}(\beta_2)}{\C Z^*_{SU(\bb n)}(\beta_1,\lambda_0^-)\C Z_{SB_1U(\bb n)}(\beta_1,\lambda_t)\C Z_{B_2}(\beta_2)}.
 \end{split}
\end{equation}
Now, using the defining porperty of the Hamiltonian of mean force [Eq.~(3) in the main text], we confirm that the term 
involving the partition functions cancels. We now sort the terms in the entropy production into terms depending on the 
final time $t$ and terms depending on the initial time $t_0^-$: 
\begin{equation}
 \begin{split}
  \Sigma(t) =&~ S[\rho_{SU(\bb n)}(t)] 
  + \mbox{tr}\left\{\rho_\text{tot}(t)\left[\ln\pi^*_{SU(\bb n)}(\beta_1,\lambda_t) 
     - \ln\pi_{SB_1U(\bb n)}(\beta_1,\lambda_t)\otimes\pi_{B_2}(\beta_2)\right]\right\} \\
  & - S[\rho_{SU(\bb n)}(t_0^-)]
  - \mbox{tr}\left\{\rho_\text{tot}(t_0^-)\left[\ln\pi^*_{SU(\bb n)}(\beta_1,\lambda_0^-) 
     - \ln\pi_{SB_1U(\bb n)}(\beta_1,\lambda_0^-)\otimes\pi_{B_2}(\beta_2)\right]\right\}.
 \end{split}
\end{equation}
Next, we use that the initial time $t_0^-$ prior to the first system-unit interaction is chosen such that 
$V_{SU(0)}(\lambda_0^-) = 0$, which implies 
$\pi^*_{SU(\bb n)}(\beta_1,\lambda_0^-) = \pi^*_S(\beta_1,\lambda_0^-)\otimes\pi_{U(\bb n)}(\beta_1)$ and 
$\pi_{SB_1U(\bb n)}(\beta_1,\lambda_0^-) = \pi_{SB_1}(\beta_1,\lambda_0^-)\otimes\pi_{U(\bb n)}(\beta_1)$. This means 
that we can replace $\pi_{U(\bb n)}(\beta_1)$ by $\rho_{U(\bb n)}(t_0^-)$, which cancels out in any case: 
\begin{equation}
 \begin{split}
  \Sigma(t) =&~ S[\rho_{SU(\bb n)}(t)] 
  + \mbox{tr}\left\{\rho_\text{tot}(t)\left[\ln\pi^*_{SU(\bb n)}(\beta_1,\lambda_t) 
     - \ln\pi_{SB_1U(\bb n)}(\beta_1,\lambda_t)\otimes\pi_{B_2}\right]\right\} \\
  & - S[\rho_{SU(\bb n)}(t_0^-)]
  - \mbox{tr}\left\{\rho_\text{tot}(t_0^-)\left[\ln\pi^*_{S}(\beta_1,\lambda_0^-)\otimes\rho_{U(\bb n)}(t_0^-) 
     - \ln\pi_{SB_1}(\beta_1,\lambda_0^-)\otimes\rho_{U(\bb n)}(t_0^-)\otimes\pi_{B_2}\right]\right\}.
 \end{split}
\end{equation}
Now, by comparing with Eq.~(\ref{eq initial state general}), we see that the very last term is nothing else than 
(minus) the initial von Neumann entropy of the universe, which is preserved during the unitary time-evolution. Hence, 
\begin{equation}
 \begin{split}
  \Sigma(t) =& -S[\rho_\text{tot}(t)] + S[\rho_{SU(\bb n)}(t)] 
  + \mbox{tr}\left\{\rho_\text{tot}(t)\left[\ln\pi^*_{SU(\bb n)}(\beta_1,\lambda_t) 
     - \ln\pi_{SB_1U(\bb n)}(\beta_1,\lambda_t)\otimes\pi_{B_2}\right]\right\} \\
  & - S[\rho_{SU(\bb n)}(t_0^-)]
  - \mbox{tr}\left\{\rho_\text{tot}(t_0^-)\ln\pi^*_{S}(\beta_1,\lambda_0^-)\otimes\rho_{U(\bb n)}(t_0^-)\right\}.
 \end{split}
\end{equation}
Next, again by using the particular form~(\ref{eq initial state general}) of the initial state, we notice that the last 
line cancels. By using the quantum relative entropy, we are hence left with 
\begin{equation}
 \Sigma(t) 
 = D\left[\rho_\text{tot}(t)\left\|\pi_{SB_1U(\bb n)}(\beta_1,\lambda_t)\otimes\pi_{B_2}\right]\right. 
 - D\left[\rho_{SU(\bb n)}(t)\left\|\pi_{SU(\bb n)}^*(\beta_1,t)\right]\right..
\end{equation}
By monotonicity of relative entropy this term is evidently positive. 

We now turn to the trajectory representation defined by a certain sequence of measurement results $\bb r_n$. They are 
obtained in the same way as for a single heat bath by measuring each unit after the interaction with the system, see the 
main text. We use the following definitions which extend our previous ones to the case of two heat baths: 
\begin{align}
 \text{stochastic work and power: } & w(\bb r_n,t) = \int_{t_0}^t ds \dot w(\bb r_n,s), ~~~ 
 \dot w(\bb r_n,t) = \lr{\frac{\partial H_\text{tot}(\lambda_t)}{\partial t}}(\bb r_n,t), \\
 \text{stochastic internal energy: } & e_{SU(\bb n)}^*(t) = \lr{H_{SU(\bb n)}^*(\lambda_t) + \beta_1\partial_{\beta_1} H_{SU(\bb n)}^*(\lambda_t) + V_{SB_2}(\lambda_t)}(\bb r_n,t), \\
 \text{stochastic measurement heat: } & q_\text{meas}(\bb r_n,t) = e_{U(\bb n)}(\bb r_n,t) - E_{U(\bb n)}(t), \\
 \text{stochastic heat flux from $B_1$: } & q_1(\bb r_n,t) = \Delta e_{SU(\bb n)}^*(\bb r_n,t) - w(\bb r_n,t) - q_2(\bb r_n,t) - q_\text{meas}(\bb r_n,t), \\
 \text{stochastic heat flux from $B_2$: } & q_2(\bb r_n,t) = -\mbox{tr}\left\{\left[\rho_{B_2}(\bb r_n,t) - \rho_{B_2}(t_0^-)\right] H_{B_2}\right\}, \\
 \text{stochastic entropy: } & s_{SU(\bb n)}(\bb r_n,t) = S[\rho_{SU(\bb n)}(\bb r_n,t)] + \beta_1^2\lr{\partial_{\beta_1} H_{SU(\bb n)}^*(\lambda_t)}(\bb r_n,t) - \ln p(\bb r_n).
\end{align}
Note that the definition of stochastic measurement heat is the same as in Eq.~(\ref{eq meas heat}) describing 
the random changes in the unit energy due to the measurement backaction. Furthermore, due to the fact that we measure 
only the units and perform no real-time feedback control, we have similar to Eq.~(14) of the main text 
\begin{equation}
 \sum_{\bb r_n}\tilde\rho_{SB_1B_2}(\bb r_n,t) = \sum_{\bb r_n}p(\bb r_n) \rho_{SB_1B_2}(\bb r_n,t) = \rho_{SB_1B_2}(t), 
\end{equation}
where $\rho_{SB_1B_2}(t) = \mbox{tr}_{U(\bb n)}\{\C U_{t,t_0}\rho_\text{tot}(t_0^-)\}$ is the unmeasured state of the 
system and baths without any measurements. This allows us to confirm the following two essential properties: 
\begin{align}
 q_2(t) &= \sum_{\bb r_n} p(\bb r_n) q_2(\bb r_n,t) = Q_2(t), \\
 q_1(t) &= \sum_{\bb r_n} p(\bb r_n) q_1(\bb r_n,t) = Q_1(t).
\end{align}
Thus, the stochastic entropy production 
$\sigma(\bb r_n,t) = \Delta s_{SU(\bb n)}(\bb r_n,t) - \beta_1 q_1(\bb r_n,t) - \beta_2 q_2(\bb r_n,t)$ 
becomes on average 
\begin{equation}
 \sum_{\bb r_n} p(\bb r_n)\sigma(\bb r_n,t) 
 = \sum_{\bb r_n} p(\bb r_n)\Delta s_{SU(\bb n)}(\bb r_n,t) - \beta_1 Q_1(t)  - \beta_2 Q_2(t).
\end{equation}
Hence, we obtain 
\begin{equation}
 \sum_{\bb r_n} p(\bb r_n)\sigma(\bb r_n,t) - \Sigma(t) = 
 \sum_{\bb r_n} p(\bb r_n) \left\{S[\rho_{SU(\bb n)}(\bb r_n,t)] - \ln p(\bb r_n)\right\} - S[\rho_{SU(\bb n)}(t)].
\end{equation}
This term is formally identical to Eq.~(17) of the main text and its positivity follows by virtue of the same arguments. 
Hence, 
\begin{equation}
 \sum_{\bb r_n} p(\bb r_n)\sigma(\bb r_n,t) \ge \Sigma(t) \ge 0.
\end{equation}

Thus, we showed how to extend quantum stochastic thermodynamics to multiple heat bath. Unfortunately, even 
if we do not drive the system-bath interaction $V_{SB_i}$, the definitions proposed here are not fully operational 
in the sense that they cannot be computed by knowing solely the state of the system and the units. To evaluate the 
heat flow $q_2$ knowledge about the state of $B_2$ is necessary. It should be emphasized, however, that this 
does not indicate a particular shortcoming of our approach. It is known that in case of multiple heat bath 
already classical stochastic thermodynamics (even in the weak coupling and Markovian regime) is not operational in 
the sense above unless additional assumptions are made. For further details see the discussion in Sec.~VII~A in 
Ref.~\cite{StrasbergPRE2019}. 

\end{widetext}

\end{document}